\documentclass[english, aps, prb, twocolumn, floatfix, superscriptaddress, 10pt]{revtex4-2}
\usepackage{hyphenat}
\usepackage[utf8]{inputenc}
\usepackage{color}
\usepackage[english]{babel}
\DeclareUnicodeCharacter{2215}{\textbackslash}
\usepackage{nicefrac}
\usepackage{multirow}
\usepackage{amsmath}
\usepackage{amssymb}
\usepackage{graphicx}
\usepackage{microtype}
\usepackage{pgfkeys}
\usepackage{acronym}
\usepackage{siunitx}
\usepackage{diagbox}
\usepackage[
    unicode=true,
    bookmarks=true,
    bookmarksnumbered=false,
    bookmarksopen=false,
    breaklinks=false,
    pdfborder={0 0 0},
    pdfborderstyle={},
    backref=false,
    colorlinks=true
]{hyperref}

\usepackage[a4paper, margin=0.75in]{geometry}

\usepackage{xcolor}
\definecolor{blue}{rgb}{0.1216, 0.4667, 0.7059}
\definecolor{orange}{rgb}{1.0000, 0.4980, 0.0549}
\definecolor{green}{rgb}{0.1725, 0.6275, 0.1725}
\definecolor{red}{rgb}{0.8392, 0.1529, 0.1569}
\definecolor{violet}{rgb}{0.5804, 0.4039, 0.7412}
\definecolor{brown}{rgb}{0.5490, 0.3373, 0.2941}
\definecolor{pink}{rgb}{0.8902, 0.4667, 0.7608}
\definecolor{gray}{rgb}{0.4980, 0.4980, 0.4980}
\definecolor{olive}{rgb}{0.7373, 0.7412, 0.1333}
\definecolor{cyan}{rgb}{0.0902, 0.7451, 0.8118}
\definecolor{change}{rgb}{0.1, 0.5, 0.00}

\usepackage[normalem]{ulem}

\newcommand{\NREP}{\ensuremath{P^{\rm NRE}}}

\def\doibasefix#110.{https://doi.org/10.}

\usepackage{siunitx}[onecolumn]
\usepackage{printlen}

\begin{document}
% \uselengthunit{in} 
% The column width is: \printlength{\textwidth}
% The column width is: \printlength{\columnwidth}

\title{Spectral shadows of a single GaAs quantum dot}
\author{Kai H{\"u}hn}
\author{Lena Klar}
\author{Fei Ding}
\affiliation{Institut f{\"u}r Festk{\"o}rperphysik, Leibniz Universit{\"a}t Hannover, Appelstra\ss e 2, 30167 Hannover, Germany}
\affiliation{Laboratory of Nano and Quantum Engineering, Leibniz Universit{\"a}t Hannover, Schneiderberg 39, 30167 Hannover, Germany}

\author{Arne Ludwig}
\author{Andreas D. Wieck}
\affiliation{Lehrstuhl für Angewandte Festkörperphysik, Ruhr-Universität Bochum, DE-44780 Bochum, Germany}

\author{Jens H{\"u}bner}
\email{jhuebner@nano.uni-hannover.de}
\author{Michael Oestreich}
\email{oest@nano.uni-hannover.de}
\affiliation{Institut f{\"u}r Festk{\"o}rperphysik, Leibniz Universit{\"a}t Hannover, Appelstra\ss e 2, 30167 Hannover, Germany}
\affiliation{Laboratory of Nano and Quantum Engineering, Leibniz Universit{\"a}t Hannover, Schneiderberg 39, 30167 Hannover, Germany}
\date{\today}

\begin{abstract}
Semiconductor quantum dots are a leading candidate for producing single and entangled photons. However, even in state of the-art devices, their performance is constrained by fluctuations in the charge state of the quantum dot and its surrounding environment. In this work, we carry out detailed time-resolved resonance fluorescence measurements on an individual charge-tunable GaAs quantum dot, providing new insight into the spectral signatures generated by a complex landscape of impurities. By varying the laser detuning, we uncover multiple Stark-shifted resonances linked to rare spectral jumps that are smaller than the homogeneous linewidth and thus are usually buried in the measurement noise. We find comparable environmentally driven Stark shifts for both the neutral exciton and the negatively charged trion transitions, whereas the positively charged and doubly negatively charged trions behave markedly differently. Our analysis quantifies the underlying impurity charge dynamics across timescales ranging from well below a millisecond up to seconds and shows that the hole population of the positively charged trion is limited by fast hole loss combined with slow hole recapture. By introducing a second, non-resonant laser, we enhance the hole occupancy by more than an order of magnitude and observe both a higher hole tunneling rate into the quantum dot and an increased hole residence time. These conclusions are corroborated by complementary spin noise spectroscopy, which provides a much higher measurement bandwidth than the time-resolved resonance fluorescence technique.
\end{abstract}

\maketitle

\section{Introduction}
Semiconductor quantum dots (QDs) provide a promising platform for single and entangled photon sources, quantum repeaters, spin- and photon-based quantum logic, and quantum light sources for quantum imaging \cite{zhai_low-noise_2020, neuwirth_quantum_2021, arakawa_progress_2020}. Compared to single atoms, individual QDs offer significantly greater flexibility in terms of experimental accessibility and the engineering of their optical properties. However, despite the tremendous progress in the growth and manufacturing processes of such QDs, several obstacles still limit their optimal use in quantum devices. One major impediment results from impurities in the surrounding solid-state matrix that lead to charge noise, resulting in stochastic shifts of the Fourier transform-limited QD transitions, blinking, and decoherence. This noise can be strongly reduced by embedding the heterostructure containing the QD layer in a PIN diode structure. However, some noise remains even in the most advanced PIN structures. This noise reveals itself in time-dependent spectral shadow resonances, which will be centrally addressed in this work, and is accompanied by charge-changing Auger recombination in the case of resonant excitation of charged exciton resonances. 

\begin{figure*}[t] % Use [tb] for top or bottom placement
    \includegraphics[]{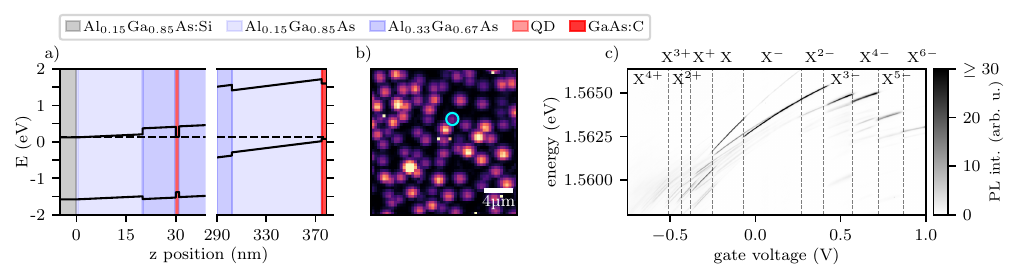}
    \caption{
        a) Schematic band structure of the sample including the QD layer located \qty{30}{\nm} from the n-contact and \SI{349}{\nm} from the p-contact. The black horizontal dashed line depicts the Fermi-level at zero bias.
        b) Photoluminescence map with the QD of interest marked by a blue circle.
        c) Photoluminescence spectra in dependence on gate voltage in a grayscale color code. The sample temperature is \qty{4}{\K}, the excitation power \qty{20}{\micro\W}, the focus diameter $\approx 1\unit{\um}$, and the excitation photon energy \qty{1.592}{\eV}.\label{fig:band_structure_pl_map}
    }
\end{figure*}

The influence of charge and spin fluctuations on quantum dot transitions has been a central topic in recent research. Studies have revealed that single-charge fluctuations at the GaAs/(AlGa)As interface in a Schottky diode can lead to significant step-like Stark shifts in the resonance of nearby (InGa)As QDs. These charge centers, which are primarily localized holes, are not randomly distributed across the 2D plane but rather form due to the strain induced by the (InGa)As lattice mismatch and the roughness of the GaAs/(AlGa)As interface. This suggests that the QD itself prompts the creation of a confined number of localization centers directly above it \cite{houel_probing_2012}.
Further investigations have been made on the temporal dynamics of charge and spin noise across frequencies ranging from \qty{0.1}{\Hz} to \qty{100}{\kHz}, showing charge noise to be particularly dominant at low frequencies, with transform-limited QD optical linewidths being achievable by operating above \qty{50}{\kHz} \cite{kuhlmann_charge_2013}.
In Ref.~\cite{hauck_locating_2014}, resonant laser spectroscopy is used to spatially pinpoint charge fluctuators within the semiconductor matrix surrounding multiple (InGa)As QDs. Their findings suggested that not only localized holes at the hetero-interface but also remaining carbon dopants in the bulk could act as significant charge-fluctuating traps in these structures. The interplay of hole tunneling between these carbon impurities and the resonances of the two-dimensional hole gas accounted for the steady-state impact of these impurity sites.
Parallel advancements have been made in stabilizing the charge environment of a single QD. Techniques such as deploying feedback using the phonon-assisted component of resonance fluorescence have successfully stabilized both the frequency and photon emission rate of the zero-phonon transition \cite{hansom_frequency_2014}.

After identifying the influence of fluctuating surface charges on the spectral diffusion of GaAs quantum dots, with a noted dependency on the QD's morphology \cite{ha_size-dependent_2015}, subsequent advances have been made in noise mitigation and characterization techniques.
Another built-in feedback loop has been developed to effectively suppress photon noise \cite{al-ashouri_photon_2019}, and single QDs have been utilized as quantum sensors to precisely measure the position and activation energy of individual shallow impurities at the nanoscale \cite{kerski_quantum_2021}.
Additionally, the optimization of a PIN diode structure has led to the integration of low-noise GaAs QDs, enhancing their suitability for quantum photonics applications \cite{zhai_low-noise_2020}.

Most of these experiments rely on resonance fluorescence (RF) spectroscopy in PIN diode structures.
Tuning the RF laser to specific QD resonances allows for selective probing of particular charge states and the real-time observation of charge dynamics, including carrier capture, tunneling events, and relaxation processes. Experiments using pulsed RF measurements on (InGa)As QDs, which are coupled to a charge reservoir through a \SI{30}{\nm} tunnel barrier, revealed, for example, electron tunneling rates in the QD $\gamma_\text{e}$ of \SI{200}{\kHz} and Auger emission rates $\gamma_\text{a}$ of \SI{500}{\kHz} \cite{kurzmann_auger_2016}. These tunneling and Auger emission rates are of interest because the homogeneous QD linewidth is broadened and the photon emission rate is quenched for high ratios of $\gamma_\text{a} / \gamma_\text{e}$. Auger rates in positively charged (InGa)As QDs have also been investigated using spin noise spectroscopy (SNS), reporting hole emission rates of \SI{2}{\MHz}, which limit the hole spin lifetime for resonant experiments \cite{wiegand_hole-capture_2018}. Recent advancements in time-resolved RF measurements have enabled optical real-time monitoring of single-electron tunneling events. In conjunction with full counting statistics evaluation methods, a new benchmark was established in the study of Auger emissions. Measurements using this method revealed additional, unassigned electron spin relaxation channels with a common rate of \SI{3}{\kHz} for (InGa)As QDs \cite{kurzmann_optical_2019}. Real-time RF measurements using two colors allowed for further studies of Auger events in a single QD \cite{lochner_real-time_2020}. An Auger emission time of \qty{1.7}{\micro\s} and electron tunneling rates of 1$\,$kHz have been reported for a tunnel barrier with a thickness of \SI{45}{\nm}. As expected, the effective Auger rate can be controlled by the probe beam power and shows a linear dependence. Other two-color experiments using low-energy photons revealed the photoeffect as a previously unnoticed non-resonant charge-loss channel for QDs~\cite{lochner_internal_2021}. Shortly after, this photoeffect was confirmed in SNS experiments in single (InGa)As QDs \cite{sterin_two-way_2023}.

In the following, we will study the carrier dynamics in and around a nearly unstrained single GaAs QD located in an optimized, low-noise PIN diode structure using time-resolved RF spectroscopy.  We will quantitatively resolve charge-induced shifts of the QD resonance, some of which are significantly smaller than the homogeneous QD transition linewidth. We will discuss the charging dynamics and the impact of non-resonant optical excitation and show how charging dynamics occurring faster than typical RF binning affect the RF spectrum.

%%%%%%%%%%%%%%%%%%%%%%%%%%%%%%%%%%%%%%%%%%%%%%%%%%%%%%%%%%%%%%%%%%%%%%%%%%%%%%%%%%
\section{Experimental details}
\begin{figure*}
    \centering
    \includegraphics[]{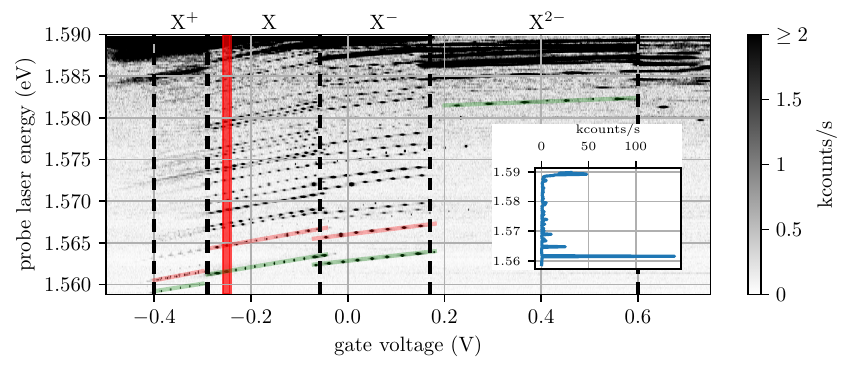}
    \caption{\label{fig:RF_E_Scan_2D}
        Resonance fluorescence measurements showing the charge plateaus from X$^+$ to X$^{2-}$, including both ground and numerous excited states. Ground states ($n=1$) are highlighted in green and the first excited states ($n=2$) in red. The grayscale color code of the photon rate is adapted so that even low photon rates are visible. The resonance plateaus appear discontinuous due to the \qty{200}{\micro\eV} step size of the laser's photon energy employed in this experiment. The inset depicts exemplarily the average photon rate measured in the regime marked in red.
    }
\end{figure*}

The high quality GaAs QDs are grown by molecular beam epitaxy using the well-developed nano-droplet technology \cite{babin_charge_2021, da_silva_gaas_2021}. The QDs are embedded in a PIN diode structure depicted in Fig.~\ref{fig:band_structure_pl_map}(a), which allows for the deterministic charging of the QD and strongly reduces environmental charge noise. The PIN structure consists of a \SI{150}{\nm} Si-doped (Al$_{0.15}$Ga$_{0.85}$)As n-contact, an undoped \SI{20}{\nm} (Al$_{0.15}$Ga$_{0.85}$)As and 10\unit{\nm} (Al$_{0.33}$Ga$_{0.67}$)As tunneling barrier, \SI{0.3}{\nm} AlAs, droplet etched GaAs QDs, an undoped \SI{274}{\nm} (Al$_{0.33}$Ga$_{0.67}$)As and \SI{75}{\nm} (Al$_{0.15}$Ga$_{0.85}$)As spacer layer, and a \SI{5}{\nm} C-doped GaAs p-contact. 
Figure~\ref{fig:band_structure_pl_map}(b) shows a spatially and spectrally resolved photoluminescence (PL) map of the sample at a temperature of \SI{4}{\kelvin} and zero bias voltage. We consistently present results from the same QD throughout this publication. The QD is marked in Fig.~\ref{fig:band_structure_pl_map}(b) by a blue circle and exhibits a single PL emission line at low excitation intensities (not displayed), which is spectrally well separated from the emission lines of the neighboring QDs that emit at higher energies. 

The PL and RF measurements are performed in a high-resolution confocal microscope setup with a focal spot diameter of $\lesssim \qty{1}{\um}$ in reflection geometry using tunable diode and Ti:Sapphire ring lasers. For PL measurements, the QDs are excited non-resonantly by a diode laser with a photon energy of \qty{1.63}{\eV}, and the detected PL of individual QDs is spectrally resolved using a triple-stage spectrometer with a maximum resolution of \qty{20}{\micro\eV} and a liquid-nitrogen-cooled CCD camera. For RF measurements, the excitation laser is suppressed using a cross-polarization extinction technique \cite{kuhlmann_dark-field_2013} without any additional wave plates, achieving a suppression of the laser stray light of about $1:10^{-6}$. In the two-color experiments, the non-resonant laser is further suppressed using a band-pass filter \footnote{We observed that the polarization extinction exhibits a significant wavelength dependence, making the additional color filter essential for the far-detuned non-resonant laser}. The photons emitted by the QD are detected by a single photon avalanche photodiode (APD)\footnote{Excelitas SPCM-AQRH-15-FC, saturation rate: \SI{37}{\MHz}} and a high-speed digitizer card \footnote{Alazartech ATS9360, which operates at a sampling rate of \SI{180}{\MHz}}, recording the arrival times of single photons from the APD. The sample is kept continuously in a closed-cycle cryostat at a temperature of 4$\,$K, allowing extended measurement durations without recurrent thermal cycling for coolant replacement. High-precision, low-temperature piezo actuators \footnote{Attocube ANSxyz100std/LT, ANPx102/RES/LT/HV, and ANPz102/RES/LT/HV} provide accurate and stable position control.

%%%%%%%%%%%%%%%%%%%%%%%%%%%%%%%%%%%%%%%%%%%%%%%%%%%%%%%%%%%%%%%%%%%%%%%%%%%%%%%%%%
\section{QD charge state characterization}
A first characterization of the QD and its static charge state is carried out by recording PL spectra as a function of the gate voltage $V_G$. Figure~\ref{fig:band_structure_pl_map}(c) shows a collection of PL spectra of the studied QD for gate voltages ranging from -0.75\,V to 1\,V.
Here, emission signatures corresponding to exciton complexes ranging from $X^{4+}$ to $X^{6-}$ are observed, with the most prominent transitions occurring for $X^+$, $X$, and $X^-$. Adjacent charge plateaus exhibit overlapping regions where the QD charge state is unstable and switches between two charge states.

A more detailed characterization of the QD and its static charge state is carried out by RF, dependent on the gate voltage ($V_G$) and the RF probe laser energy. Figure~\ref{fig:RF_E_Scan_2D} depicts the two-dimensional RF measurement revealing distinct charge plateaus for the positively charged trion ($X^+$), the neutral exciton ($X$), the negatively charged trion ($X^-$), and the doubly negatively charged exciton ($X^{2-}$) transition. The assignment of the transitions is confirmed by the polarization and $V_G$ dependent PL measurements shown in Fig.~\ref{fig:band_structure_pl_map}(c), which yield a fine structure splitting of the neutral exciton of \qty{1.6(1)}{\micro\eV}
and accordingly shows no splitting for the negatively charged trion transition. A comparison of the PL spectra with the corresponding RF spectra in Fig.~\ref{fig:RF_E_Scan_2D} reveals a significant Stokes shift for $X^{2-}$. This occurs because RF probes the $n=2$ absorption transition, while the PL arises primarily from the $n=1$ transition due to the rapid relaxation of the excited hole to $n=1$.

\sisetup{per-mode = single-symbol} %
In the following, we focus on the X$^+$, X, X$^-$, and X$^{2-}$ ground state transitions. The QD transition energies in Fig.~\ref{fig:RF_E_Scan_2D} show a clear dependence on the gate voltage $V_G$ due to the quantum confined Stark effect. In general, the exciton transition energy, $E_X$, exhibits a linear and quadratic dependence on the local electric field amplitude in the $z$-direction, $F_z$, according to~\cite{xu_manipulating_2008, fry_inverted_2000}:
\begin{equation}
    \label{eq:Stark_shift}
    E_X(F)=E_{x}(0) - p_z F_z + \beta F_z^2,
\end{equation}
where $E_{X}(0)$ is the exciton transition energy in the absence of an externally applied electric field, $p_z$ is the accessible part of the permanent dipole moment arising from the intrinsic spatial separation of the electron and hole in the anisotropic QD, and $\beta$ is the polarizability of the electron-hole pair. A fit to the measured exciton ground state transition energy versus $V_G$ yields $p_z = \qty{9.1(0)}{\milli\eV} \cdot (\qty{380}{\nm\per\V})$ and $\beta = \qty{-2.7(1)}{\milli\eV} \cdot (\qty{380}{\nm\per\V})^2$. For the charged exciton states, the additional carrier in the QD partially screens the electric field. This screening is negligible in the case of $X^+$ since the holes are significantly more localized than the electrons. In fact, the measured dipole moment $p_z$ is identical for $X$ and $X^+$ within the measurement uncertainty. In contrast, $p_z$ decreases significantly for $X^-$ and $X^{2-}$ by \SI{26}{\percent} and \SI{54}{\percent}, respectively, while the respective polarizability $\beta$ decreases by \SI{13}{\percent} and increases by \SI{57}{\percent}. For $X^+$, $\beta$ decreases by \SI{63}{\percent}. We want to point out that these changes in $\beta$ strongly depend on the QD morphology and charge occupation, making general theoretical predictions difficult.

\begin{figure*}[tb]
    \centering
    \includegraphics[]{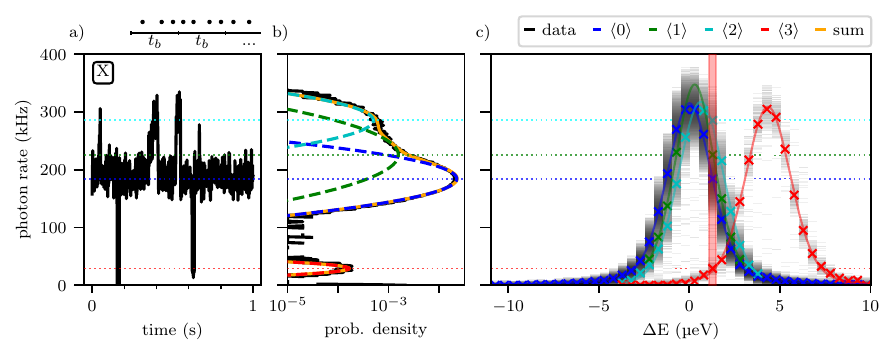}
    \caption{
        (a) Excerpt of a typical telegraph-like signal recorded on the neutral exciton transition $X$ for a detuning of  $\Delta E = 1.29\,\unit{\micro\eV}$. Generally, the photon emission rate remains steady at approximately \qty{200}{\kHz}, but occasionally it abruptly changes to different rates.
        The top panel schematically depicts the single photons (dots) and the time intervals $t_b$  in which the fluctuating photon emission rate is averaged, i.e., binned into one point, which enters the probability analysis of the telegraph photon rates.
        (b) Relative probabilities of the telegraph photon rates for a total recording time of \SI{100}{\second} and a binning time of $t_b=\qty{1}{\ms}$.
        The four clearly distinguishable peaks are labeled as states $\langle 0\rangle\ldots\langle 3\rangle$: The dominant resonance $\langle 0\rangle$, two weakly shifted resonances $\langle 1\rangle$ and $\langle 2\rangle$, and the strongly shifted resonance $\langle 3\rangle$. The relative occurrences of the states evaluate to 90.9\,\%, 6.3\,\%, 2\,\%, and 0.7\%, respectively. A barely visible peak at 1\,kHz corresponds to a non-bright state.
        (c) Measured detuning dependence of the photon rate. The spectral position, where the data shown in (a) and (b) is recorded, is marked in red at a detuning of $\Delta E=\qty{1.29}{\micro\eV}$. The dotted horizontal lines across the figures mark the center position of the peaks determined in (b).
    }
    \label{fig:tel_rel_ons}
\end{figure*}

%%%%%%%%%%%%%%%%%%%%%%%%%%%%%%%%%%%%%%%%%%%%%%%%%%%%%%%%%%%%%%%%%%%%%%%%%%%%%%%%%%%%%%%%%%%%%%%%%%%%%
\section{QD charge dynamics}
In the following, we will use the quantum confined Stark shift of the QD as a probe for the local electric field in order to measure the charging dynamics of the QD and its surroundings.

\subsection{Quasi static dynamics}
First, we will study the average quasi static distributions of the neutral exciton transition $X$.
Figure~\ref{fig:tel_rel_ons}(a) shows a typical telegraph-like signal recorded on the neutral $X$ transition for a detuning of the RF laser from the transition maximum of the main state of $\Delta E=\qty{1.29}{\micro\eV}$, which corresponds to about half the width at half maximum of the transition. For the data shown, the single-photon events detected by the APD are grouped into non-overlapping time intervals with a binning period $t_b$ of \qty{1}{\ms} as indicated in the inset above. At this specific detuning, the RF photon rate depicted in Fig.~\ref{fig:tel_rel_ons}(a) remains steady at approximately \qty{200}{\kHz} for most of the time, but it occasionally changes for a brief period.
These changes result from a slight shift in the QD resonance, for example, due to a charge change in the QD surroundings. In contrast, a change in the charge state of the QD itself, i.e., from $X$ to $X^+$, results in a significantly larger jump in the transition energy, as can be seen from Fig.~\ref{fig:RF_E_Scan_2D}, which exceeds the QD linewidth by far and results in a reduction of the RF signal to effectively zero. Figure~\ref{fig:tel_rel_ons}(b) shows the same measurement as Fig.~\ref{fig:tel_rel_ons}(a) but depicts the relative probability of each photon rate in black for a total recording time of \qty{100}{\s}.
The probability density of the distinct photon rates shown in Fig.~\ref{fig:tel_rel_ons}(b), extracted with $t_b=\SI{1}{\ms}$, can be accurately fitted by Gaussian functions. Each Gaussian center value corresponds to the shot noise level for the respective photon count rates, which are depicted as horizontal dashed lines.
The good agreement between this model and the data shows that the measurement is shot noise limited.
An increase of $t_b$ reduces the relative shot noise level but degrades the time resolution and vise versa. The peak observed at a very low photon rate of $\lessapprox1$~kHz, with a photon emission rate probability of $~2\cdot 10^{-3}$, corresponds to a far off-resonant state, which does not emit any photons because it is no longer excited by the probing laser.
In this case, the detected photons result solely from leakage due to limited polarization extinction and from the intrinsic APD background count rate of about 50\,Hz. The area of each curve is proportional to the occurrence probability of the corresponding scenario (78.6\,\%, 19.9\,\%, and 1.5\,\%) \footnote{Strictly speaking, these values contain a very small systematic error, as both photon leakage and dark counts ($\lessapprox \SI{1}{\kHz}$) are always present.}.

We want to point out that each of these resonances results from a specific charging configuration of the surrounding environment of the QD, labeled $\langle 0\rangle$ to $\langle 3\rangle$. The center positions of the four Gaussian curves specify the respective brightnesses of the four configurations for a given $\Delta E$ and are marked in Fig.~\ref{fig:tel_rel_ons}(c) by crosses.

Next, we repeatedly conduct the experiment introduced in Fig.~\ref{fig:tel_rel_ons}(a,b), varying the detuning energy $\Delta E$ from \qty{-10}{\micro\eV} to \qty[print-implicit-plus]{10}{\micro\eV} of the laser with respect to the maximum of the dominant $X$ resonance.
This approach provides much more detailed insight into the relevant number of impurity configurations in the QD surroundings and their charging dynamics.
Figure~\ref{fig:tel_rel_ons}(c) shows the measured RF photon rate, in a grayscale color code, as a function of $\Delta E$. The data shown in Fig.~\ref{fig:tel_rel_ons}(b) is marked by a red vertical shaded region in Fig.~\ref{fig:tel_rel_ons}(c). The complete measurement set reveals that this specific detuning corresponds approximately to the half width at half maximum of the optical transition and is thus much more sensitive to subtle spectral shifts of the QD transition energy.
The characteristic detuning dependent average photon rates are depicted as crosses in Fig.~\ref{fig:tel_rel_ons}(c) and fitted by four pseudo-Voigt functions, represented by colored lines, which yield the spectral positions of the effectively unshifted and shifted resonances. The analysis clearly shows the four different charge configurations, denoted as RF QD shadows. Three of them are very close in energy, and one is shifted from the dominant state by about \qty{4.35+-0.01}{\micro\eV}. We emphasize that the likelihoods of the charge configurations $\langle 1\rangle$ and $\langle 2\rangle$ are more than one order of magnitude lower compared to state $\langle 0\rangle$, and that the respective energy shifts from the dominant state $\langle 0\rangle$ are only \qty{0.30+-0.01}{\micro\eV} and \qty{0.64+-0.02}{\micro\eV}, which are significantly smaller than the linewidth of \qty{2.95+-0.02}{\micro\eV} of the QD resonance. In fact, our analysis clearly distinguishes sparsely occupied shadow resonances, which are usually obscured by the dominant resonance in other experiments.

% @Jens: Das sind die Differenzen zwischen dem tatsächlichen sigma und dem shotnoise-limitiertem Sigma. G1 und G2 sind demnach im Prinzip shotnoise limited, G0 und G3 nur fast.  
 % G0 diff = 2.65159411
 % G1 diff = -0.22591028
 % G2 diff = -0.51955782
 % G3 diff = 9.80567211

% @Jens: Linienbreiten mit Unsicherheiten, alles in eV
% fwhm state 0 (blue): 2.947630328509207e-06 \pm 2.1563918990750087e-08
% fwhm state 1 (green): 2.625378461697248e-06 \pm 1.278535535469786e-07
% fwhm state 2 (cyan): 2.944034204579491e-06 \pm 6.951282958920955e-08
% fwhm state 3 (red): 2.9731553670320015e-06 \pm 3.0945322925668555e-08

\begin{figure}[tb]
    \centering
    \includegraphics[]{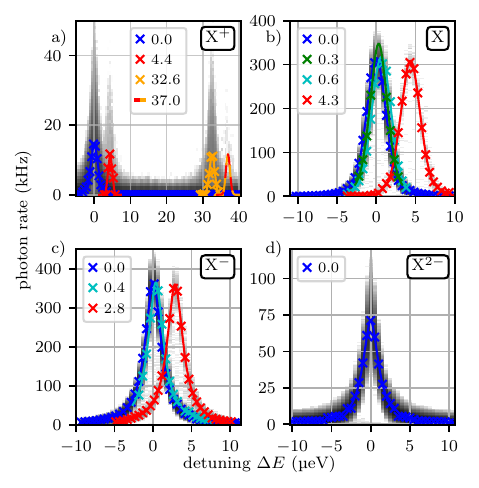}
    \caption{\label{fig:timetag_det_dep_X+_X_X-_X2-}
        Probability densities as grayscale color code in dependence on the detuning $\Delta E$ for different QD charge states. The solid lines are fits to the observed resonances using pseudo-Voigt functions. The dominant transition is marked in dark blue and defines $\Delta E = 0$ with a precision of a few \unit{\nano\eV} (see also Tab.~\ref{tab:x+_shifts}). The legends indicate the shifts of the resonances in \unit{\micro\eV}.
    }
\end{figure}

The occupation probability of the various impurity states and the screening of $F_z$ are expected to depend on both the charge occupation of the QD and the chemical potential. Figure~\ref{fig:timetag_det_dep_X+_X_X-_X2-} compares the results obtained above for measurements on $X$ with $X^+$, $X^-$, and $X^{2-}$. In each case, the dominant resonance, marked in dark blue, occurs at the lowest energy. This corresponds to the highest electric field at the QD position. Since the relevant impurities are most likely Si impurities in the vicinity of the n-contact \cite{mar_bias-controlled_2011}, this lowest energy corresponds to the scenario in which all relevant Si impurities in the unintentionally doped (AlGa)As spacer between the n-contact and the QD are each occupied by an electron. In principle, charge fluctuations from neighboring QDs could also influence the observed charging dynamics, but this scenario can be excluded with high probability based on an analysis of the shadow resonances as a function of gate voltage. According to our PL measurements, neighboring QDs have a different ground state energy and therefore likely switch at a different $V_G$; however, the number and energy detuning of all observed shadow resonances of the investigated QD remain unchanged within the studied QD charge plateaus. Furthermore, optical excitation of neighboring QDs can be ruled out due to their spatial and spectral separation \footnote{The neighboring QDs have a significantly larger transition energy, which probably inhibits charge changes at moderate $V_G$.}.

The gate voltage dependent resonance fluorescence data depicted in Fig.~\ref{fig:RF_E_Scan_2D} show that mainly the signatures of the resonances of $X$ and $X^-$ occur. Taking this into consideration, Fig.~\ref{fig:timetag_det_dep_X+_X_X-_X2-} confirms that the average configuration of active Si impurities is identical for both $X$ and $X^-$, where \emph{active} in this case means that the charge state changes frequently during a measurement period. In contrast, no active Si impurities are observed for $X^{2-}$, suggesting that the applied electric field is strong enough to effectively suppress the charge dynamics in the surrounding Si impurities involved. For $X^+$, on the other hand, the spectra shown in Fig.~\ref{fig:timetag_det_dep_X+_X_X-_X2-}(a) reveal a doublet structure, featuring a small shift of \qty{4.39+-0.09}{\micro\eV} and a larger shift of \qty{32.59+-0.02}{\micro\eV}.
\begin{figure}[tb]
    \centering
    \includegraphics[]{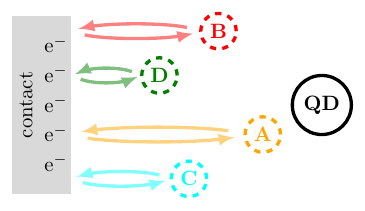}
    \caption{\label{fig:impurity_configuration}
        Schematic model of the active Si donor sites. Four distinct impurity sites are located at varying distances from the QD, each capable of trapping a single electron. Charge fluctuations of these sites induce local electric field changes, leading to characteristic shifts in the QD resonance energy.
    }
\end{figure}
All four resolvable measured configurations in Fig.~\ref{fig:timetag_det_dep_X+_X_X-_X2-}(a) to (d) can be consistently explained by effectively four distinct Si impurity sites, each situated at unique distances from the QD and capable of trapping a single electron. Figure~\ref{fig:impurity_configuration} schematically illustrates the spatial configuration of these impurity sites within the sample structure, labeled A, B, C, and D according to their increasing distance from the QD.
In order to attribute the shadow resonances to a combination of the charge states of the four identified impurities, the following notion is introduced: \textquotedblleft $-$\textquotedblright\, indicating that the impurity charge state is constant over time, and $e$ and $0$ denote that the respective Si impurity can change its charge state and is currently charged ($e$) or not charged $(0)$ with an electron. Unresolvable charge configurations are denoted by \textquotedblleft $\sim$ \textquotedblright. Also, due to the Coulomb interaction between the QD and the impurity sites, not all impurity charge states manifest for a given charge in the QD. The attributions of the configurations, together with the corresponding shifts, are listed in Tab.~\ref{tab:x+_shifts}.
For example, for $X^+$ shown in Fig.~\ref{fig:timetag_det_dep_X+_X_X-_X2-}(a), a doubled structure with a small splitting is observed, which reappears with a large shift of about \qty{33}{\micro\eV}. This large shift can be attributed to a charge change of an impurity very close to the QD, which is labeled A. The small shift leading to the doublet structure is attributed to charge changes in an impurity located at a slightly larger distance, labeled as B in this case. Further impurity charge states cannot be resolved for this resonance.
% \commentLK{Vielleicht ist er ratsam, hier dann auch den shift durch die impurity B zu erklären und die ganzen Kombinationen für X+ einmal durchzuexerzieren, wie der Reviewer vorgeschlagen hat? Dann wäre auch der Übergang zu X und X- mit "Kein großer shift wie bei X+ --> kein A" etwas klarer, und man könnte noch besser hervorheben, dass der kleine shift bei X+ von B auch bei X und X- existiert, und außerdem noch weitere impurities C und D auflösbar sind, was bei X+ nicht möglich war. Wahrscheinlich wäre es auch hilfreich, wenn die Tabelle I nahe an diesem Absatz, bestmöglich auf der gleichen Seite platziert wäre, damit man sie direkt vor Augen hat, wenn man den Absatz liest, und die Konfigurationen sehr viel besser nachvollziehen kann.}
For a charge configuration of the QD corresponding to $X$ and $X^-$, the closest Si impurity $A$ becomes inactive, and a combination of the three impurities B, C, and D, with decreasing impact onto the resonance shift, is responsible for the observed spectrum, as any combination of only two impurities does not reproduce the experimentally observed energy shifts. We want to point out that the experiment does not show all possible combinations, suggesting that the impurities interact quite strongly with one another. For $X^{2-}$, either no additional charge trap configuration beyond the main state is active, or it is unresolvable.
\begin{table}[tb]
\caption{Here, $-$, $e$, or $0$, respectively, indicate whether the impurity charge state is constantly inactive over time, or that the respective Si impurity can change its charge state and is currently charged ($e$) or not charged $(0)$ with an electron. By $\sim$, we denote an either unresolvable trap configuration or shift associated to the impurity site configuration, respectively.}
\label{tab:x+_shifts}
\centering
\begin{tabular}{|c|l|c|c|c|c|}\hline
 & \diagbox[width=\dimexpr \textwidth/8+2\tabcolsep\relax, height=1cm]{ shift }{impurity\\site} & A & B & C & D \\ \hline
\multirow{4}*{X$^+$}     & \qty{0.000+-0.009}{\micro\eV}    & $e$ & $e$ & $\sim$ & $\sim$ \\ \cline{2-6}
                         & \qty{4.39+-0.09}{\micro\eV}      & $e$ & $0$ & $\sim$ & $\sim$ \\ \cline{2-6}
                         & \qty{32.59+-0.02}{\micro\eV}     & $0$ & $e$ & $\sim$ & $\sim$ \\ \cline{2-6}
                         & \qty{36.98+-0.09}{\micro\eV}     & $0$ & $0$ & $\sim$ & $\sim$ \\ \hline

 \multirow{4}*{X$^0$}    & \qty{0.000+-0.007}{\micro\eV}    & $-$ & $e$ & $e$ & $e$ \\ \cline{2-6}
                         & \qty{0.30+-0.01}{\micro\eV}      & $-$ & $e$ & $e$ & $0$ \\ \cline{2-6}
                         & \qty{0.64+-0.02}{\micro\eV}      & $-$ & $e$ & $0$ & $e$ \\ \cline{2-6}
                         & \qty{4.35+-0.01}{\micro\eV}      & $-$ & $0$ & $e$ & $e$ \\ \hline

  \multirow{4}*{X$^-$}  & \qty{0.000+-0.004}{\micro\eV}     & $-$ & $e$ & $e$ & $e$ \\ \cline{2-6}
                        & $\sim$                            & $-$ & $e$ & $e$ & $0$ \\ \cline{2-6}
                        & \qty{0.354+-0.009}{\micro\eV}     & $-$ & $e$ & $0$ & $e$ \\ \cline{2-6}
                        & \qty{2.846+-0.006}{\micro\eV}     & $-$ & $0$ & $e$ & $e$ \\ \hline

  {X$^{2-}$}  & \qty{0.00+-0.01}{\micro\eV}    & $-$ & $-$ & $\sim$ & $\sim$ \\ \hline

\end{tabular}
\end{table}

\begin{figure}[tb]
    \centering
    \includegraphics[]{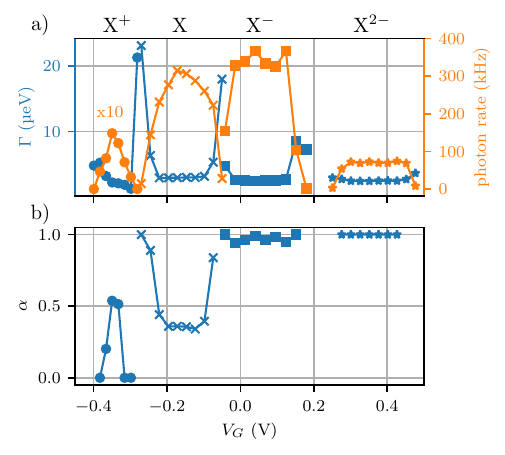}
    \caption{\label{fig:X+_X0_Xm_params_NoRL}
        (a) Photon rates and linewidth $\Gamma$, and (b) fraction parameter $\alpha$ as a function of $V_G$ for the dominant state 0, derived from pseudo-Voigt fits for X$^+$ to X$^{2-}$. The brightness of X$^+$ is multiplied by a factor of 10 for clarity. The solid lines serve as guides to the eye.
    }
\end{figure}

Next, we measure the influence of the fluctuating impurities on the dominant QD transition $X^+, X,X^-$ and $X^{2-}$ while tuning $V_G$. Figure~\ref{fig:X+_X0_Xm_params_NoRL}(a) depicts an analysis of the photon rate and width of the dominant RF transition as a function of the gate voltage, providing insight into the stability and broadening mechanisms of the different QD charge states. The photon rate of the $X^+$ state is more than an order of magnitude lower compared to the $X$ and $X^-$ states and exhibits a peak-like form, which indicates that the average occupation probability of the QD by a hole is relatively low. In fact, based on the band structure shown in Fig.~\ref{fig:band_structure_pl_map}(a), it is astonishing that this transition can be observed at all in RF at this gate voltage and temperature, and the origin of a gate voltage induced hole in the QD remains unclear. Unintentional carbon background doping and/or photo-induced carriers might play some role, but these assumptions could neither be confirmed nor rejected. Further studies concerning the $X^+$ transition through two-color, non-resonant optical excitation are presented below in more detail. For $X$ and $X^-$, the maximal RF photon rates are about equal at their respective charge plateaus, and the slightly higher maximal photon rate of $X^-$ probably results mainly from its slightly narrower linewidth. Additionally, the photon rate of $X$ also features a peak-like form, indicating that the neutral charge state is more fragile compared to the negatively charged states. The photon rate of $X^{2-}$ is by a factor of about six lower compared to $X^-$, which reflects the significantly lower oscillator strength of the $n = 2$ bound exciton state. The widths of $X^{2-}$ and $X^-$ are about equal since they are both dominated by the radiative lifetime of the excitonic $n = 1$ transition, i.e., the hole from the optically excited $X^{2-}$ transition relaxes rapidly from $n = 2$ to $n = 1$. In the transition regimes from one charge state to another, the respective linewidth increases, and the photon rate drops drastically due to fast switching between two states. The sum of the two respective photon rates does not equal the average of the respective count rates at their peaks since the fast switching leads to strong broadening and reduction of the amplitude of the respective transitions, and thereby to a reduced RF photon rate at a specific probing energy.

Figure~\ref{fig:X+_X0_Xm_params_NoRL}(b) shows the fraction parameter $\alpha$ from the pseudo-Voigt fit of the dominant transition as a function of $V_G$.
The parameter $\alpha$ determines the proportion of Gaussian and Lorentzian contributions, with $\alpha=0$ indicating a purely Gaussian lineshape and $\alpha=1$ indicating a purely Lorentzian lineshape.
Interestingly, $X^-$ and $X^{2-}$ exhibit, in good approximation, a purely Lorentzian profile, whereas $X$ shows a mixture of Gaussian and Lorentzian.
This suggests that, in this kind of low-noise PIN structure, both $X^-$ and $X^{2-}$ have the potential to outperform $X$ in terms of achieving transform-limited single-photon emission.
The lineshape of $X$ becomes more Lorentzian, and $\Gamma$ larger in the transition regions from $X$ towards $X^+$ and $X^-$, which is consistent with faster switching times. However, this regime is less interesting for single photon sources.
 
% ---------------------------------------------------------------------------------------
\subsection{Switching time analysis on \texorpdfstring{$X$}{X}}\label{sec:switching_time_analysis}

In the following, we will first address details concerning the conventional switching time analysis of the telegraph signal, which is particularly effective for slow switching times.
To this end, we select the relatively bright $X$ transition and a quasi-resonant excitation detuning with $\Delta E \approx \qty{0}{\micro\eV}$, which effectively yields only a bright and a dim photon stream. 
The chosen binning time $t_b=\qty{1}{\ms}$ of the continuous stream of photons provides the optimal balance between time resolution and signal noise for the given QD dynamics. 

\begin{figure}[bh]
    \includegraphics[]{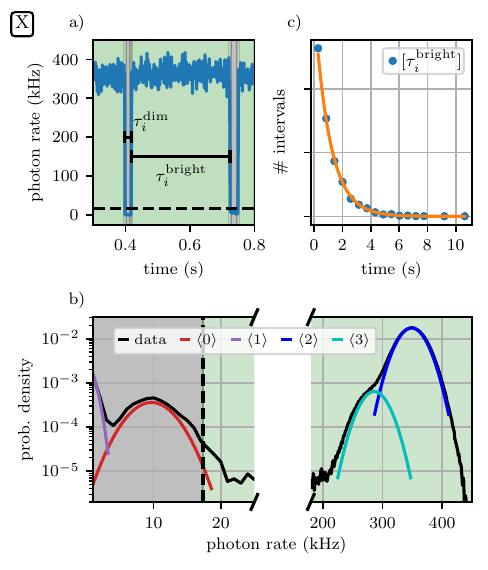}
    \caption{\label{fig:rates_single_result}
        (a) Snippet of a telegraph signal showing two switching events. The threshold (black dashed line) divides the signal into bright and dim intervals with random switching times $\tau^\text{bright}_i$ and $\tau^\text{dim}_i$. The index $i$ denotes the counter for the collected randomly occurring switching time intervals.
        (b) Exemplary probability density of $X$ for quasi-resonant excitation including all visible shadow resonances, each fitted by a Gaussian. Again, the threshold (black dashed line) separates the probability density into a bright and dim region.
        (c) Histogram of the measured $[\tau^\text{bright}_i]$ and a fit by an exponential decay (orange curve) revealing the characteristic switching time $\tau^\text{bright} = \qty{1.1(0)}{\second}$ and $\tau^\text{dim} = \qty{0.008(8)}{\second}$ (not shown). 
    }
\end{figure}

For slow switching dynamics from one photon rate level to another, with timescales significantly exceeding $t_b$, a switch is clearly resolved by the telegraph signal (as shown in Fig.~\ref{fig:rates_single_result}(a)).
In this case, the division of the telegraph signal into bright (green) and dim (gray) regions is trivial, and a straightforward switching-time analysis can be applied.
To ensure a consistent evaluation, we define a threshold photon rate, indicated in Fig.~\ref{fig:rates_single_result}(a) and (b) as a dashed black line, to distinguish between the bright and dim regions.
Generally, the bright and dim regions consist of multiple peaks, each representing a QD RF shadow resonance. We consistently set the threshold at the 99\% upper confidence interval of the peak with the highest photon emission rate in the dim region. With this approach, the varying bright and dim time constants $\tau^\text{bright}$ and $\tau^\text{dim}$ can be unambiguously collected for the total measurement time and evaluated in a histogram, as shown by the blue dots in Fig.~\ref{fig:rates_single_result}(c). The data obtained in this way is perfectly fitted by a single exponential decay.
Figure~\ref{fig:rates_single_result}(c) shows an example of such a fit, which, in this case, yields the characteristic lifetime denoted by $\tau^\text{bright}_0$ of the joint bright region given by the states $\langle 0\rangle$ and $\langle 1\rangle$, respectively.

In the case of rapid switching dynamics on the timescale of $t_b$, the telegraph signal cannot adequately resolve individual switching events, and a meaningful division into bright and dim intervals becomes impossible.
In fact, the probability density of such a telegraph signal is characterized by a prominent high background, as shown in Fig.~\ref{fig:X+_ABE_sat}(c), which originates from switching events that are shorter than the time resolution.
\subsection{Non resonant excitation case for \texorpdfstring{$X^+$}{X+}}

In the PIN diode structure, electrons tunnel efficiently between the n-contact and the QD, and the number of electrons in the QD is easily adjusted by $V_G$. The tunneling of holes out of the QD can also be modified by $V_G$, but the tunneling of holes into the QD is much slower, and the complexity of their dynamics exceeds the schematic band structure depicted in Fig.~\ref{fig:band_structure_pl_map}(a). Therefore, to study $X^+$ in greater detail, we employ an additional weak non-resonant laser excitation (NRE) with a photon energy of \qty{1.61}{\eV}. This energy is below the bandgap of the enclosing (AlGa)As barriers and excites additional electrons and holes non-resonantly directly in the QD. With an applied $V_G$ corresponding to $X^+$, lost holes are efficiently replenished, while excess electrons and holes rapidly tunnel out of the QD, making $X^+$ in the time average the most probable state. 

\begin{figure}[th]
    \includegraphics[width=\columnwidth]{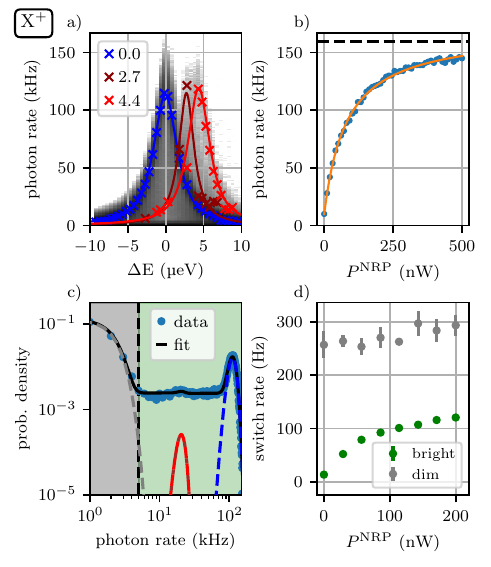}
    \caption{\label{fig:X+_ABE_sat}
        (a) Measured photon rate probability (grayscale coded) in dependence on the detuning $\Delta E$ on the positively charged trion resonance X$^+$ under above band-gap excitation with $\NREP = \qty{222}{\nano\W}$. An additional resonance appears at \qty{2.7+-0.1}{\micro\eV}.
        (b) Measured increase of the RF photon rate with increasing above band-gap excitation for a constant probe laser intensity of \qty{200}{\nano\W} (blue dots) and $\Delta E=\qty{-0.3}{\micro\eV}$. The solid orange line depicts a fit according to Eq.~\ref{eq:saturation}.
        (c) Probability density of X$^+$ measured at $\NREP = \qty{222}{\nano\W}$ and a detuning of $\Delta E=\qty{0.2}{\micro\eV}$. The probability density is fitted by three Gaussian functions along with a phenomenological background between the peaks induced by NRE. The threshold (vertical dashed line) divides the probability density into a dim (gray) and a bright (green) region.
        (d) Switching rates $\gamma^\text{dim}=1/\tau^\text{dim}$ and $\gamma^\text{bright}=1/\tau^\text{bright}$ in dependence of \NREP.
    }
\end{figure}

% \begin{figure*}[th]
% \includegraphics[]{figures/X+_ABE_sat.pdf}
%     \caption{\label{fig:X+_ABE_sat}
%         (a) Measured increase of the RF photon rate at X$^+$ with increasing above band-gap excitation for a constant probe laser intensity of \qty{200}{\nano\W} (blue dots). The solid orange line depicts a fit according to Eq.~\ref{eq:saturation}.
%         (b) Occurrence of an additional resonance at $2.7\,\mu$eV at finite NRE.
%         (c) Measured relative probabilities of the photon rates (blue dots) and simulation with a straightforward model (red line). The significant ``background'' level of approximately $2\cdot 10^{-3}$ is shown in (b) by the grayscale color code, which fills the blue fit and offers insight into the rapid charge dynamics. (b) and (c) are measured at $\NREP = \qty{222+-1}{\nano\W}$ and \commentJH{@Kai: Das ist eine unphysikalische Unsicherheit}$V_G=-0.36601(0)$\,V.
%         \\ \commentJH{@Lena: Ich möchte Fig. 8 und 9 zusammenfassen, d.h. eff. ersetzt Fig. 9 Fig8c}
%     }
% \end{figure*}

Figure~\ref{fig:X+_ABE_sat}(a) depicts the resulting detuning dependence for $X^+$ at an intermediate $\NREP = \qty{222+-1}{\nano\W}$, which shows, in addition to an order of magnitude higher photon rate, two significant differences compared to no NRE.
First, the NRE activates an additional weak shadow resonance; and second, the Gaussian peaks only describe the most probable photon count rates at a given $\Delta E$, wherein a significant background arises at all lower photon rates, which is visible in the plot as a gray shaded background. 
Furthermore, Fig.~\ref{fig:X+_ABE_sat}(b) depicts the increase of the maximum RF photon rate of the $X^+$ transition for the full range of the available NRE power \NREP. The excitation in this case is quasi-resonant with $\Delta E = \qty{-0.3}{\micro\eV}$ for the dominant state ${\langle0\rangle}$. Due to the higher occupation probability by a hole, the $X^+$ photon rate increases drastically with \NREP, showing a typical saturation behavior at high values. The solid orange line represents a fit using the saturation model 
\begin{equation}
    \label{eq:saturation}
    \gamma_{\langle0\rangle}^{ph}(\NREP) = \frac{\gamma_{\langle0\rangle}^{ph,max} \cdot \NREP}{\NREP_{sat} + \NREP} + \gamma_{\langle0\rangle}^{ph}(0)
\end{equation}
with a peak photon rate $\gamma_0^{ph,max} = \qty{150+-9}{\kHz}$, a saturation power $\NREP_{sat}$ = \qty{81+-1}{\nano\W}, and a photon rate at zero NRE of $\gamma_{\langle0\rangle}^{ph}(0) = \qty{9+-1}{\kHz}$. 

Figure~\ref{fig:X+_ABE_sat}(c) presents, for $X^+$, the probability density for non-resonant excitation at a \NREP\ of $\qty{222}{\nano\W}$ and a detuning of $\Delta E=\qty{0.2}{\micro\eV}$. At this detuning, the two shadow resonances of the $X^+$ transition, which are visible in Fig.~\ref{fig:X+_ABE_sat}(a), cannot be clearly resolved, allowing for effectively only three peaks, including a leakage contribution, to be unambiguously fitted by three Gaussian functions.
A phenomenological background accounts for the offset between the peaks, resulting from the fast switching dynamics. The threshold (indicated by the black dashed line) once again divides the probability density into a bright and a dim region. This threshold enters the analysis of the switching times in a manner similar to the evaluation introduced in Sec.~\ref{sec:switching_time_analysis}.
Figure~\ref{fig:X+_ABE_sat}(d) shows the extracted switching rates corresponding to these bright and dim intervals. Both rates increase with NRE power, although the bright state rate remains consistently lower than the dim-state rate, indicating that the QD tends to spend more time in the bright state. This is confirmed by spin noise spectroscopy (SNS) measurements, which have been carried out on the identical $X^+$ transition and are presented below.

\begin{figure}[tb]
    \includegraphics[]{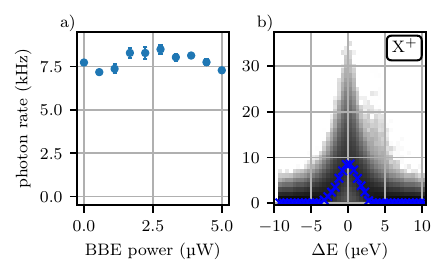}
    \caption{\label{fig:BBE}
        (a) Most probable RF photon rate at the center of the dominant $X^+$ transition versus below bandgap excitation power. (b) Grayscale color plot of the photon rate probability in dependence on detuning. The blue solid line depicts a pseudo-Voigt fit.
    }
\end{figure}

In order to exclude residual absorption in the surrounding semiconductor matrix, which could potentially supply surcharge holes, we conducted measurements of the RF photon rate of the dominant $X^+$ transition in dependence on below bandgap excitation (BBE) with a laser photon energy of \qty{1.53}{\eV}, as shown in Fig.~\ref{fig:BBE}. The laser photon energy is well below the $X^+$ QD transition and the (AlGa)As bandgaps, effectively excluding the possibility of band-to-band single photon excitation by BBE. In principle, two-photon absorption processes might be feasible and could create holes in or close to the QD. However, despite the BBE power reaching up to $\qty{5}{\micro\W}$, the measured RF photon rate remains constant, indicating that BBE does not have a measurable impact on the charging of the $X^+$ transition.

% ---------------------------------------------------------------------------------------
\section{Spin Noise Measurements\label{sec:app_SNS}}

Spin noise spectroscopy (SNS) is a versatile tool for investigating the spin and charge dynamics in semiconductors, down to the level of a single QD \cite{dahbashi_optical_2014, wiegand_hole-capture_2018, wiegand_spin_2018, sterin_two-way_2023}. Spin and charge fluctuations are imprinted onto the linear polarization axis of the probe beam and are measured using a polarization bridge and a balanced photodetector \cite{hubner_rise_2014, glazov_linear_2015}. The detector used here features a bandwidth of 220\,kHz, which increases the time resolution of the experiments from the millisecond to the microsecond regime, compared to the RF measurements presented above. The probe laser intensity was increased to $\qty{2}{\micro\W}$ for the SNS measurements to counter the reduced signal-to-noise ratio compared to the RF measurements, which utilize single photon avalanche photodiodes.
\begin{figure}[tb]
    \includegraphics[]{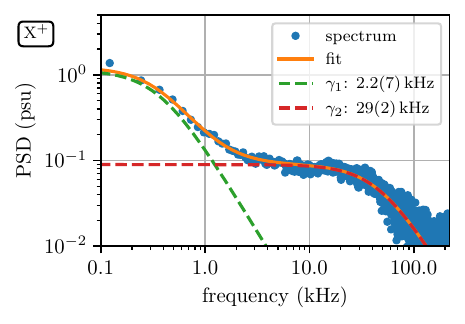}
    \caption{\label{fig:SN_typical_LL}
        Typical SN spectrum for a quasi-resonant probe laser with a detuning of $\Delta E =\qty{0.64}{\micro\eV}$ and a probe power of $\qty{2}{\micro\W}$. The power spectral density (PSD) is plotted in power spectral units (psu). The combination of two Lorentzian contributions describes the spectrum in excellent agreement. The first component (green) is identified with the Si donor charge dynamics, while the second contribution (red) is identified with the fast hole occupancy dynamics of the QD.
    }
\end{figure}
\begin{figure}[t]
    \includegraphics[width=\columnwidth]{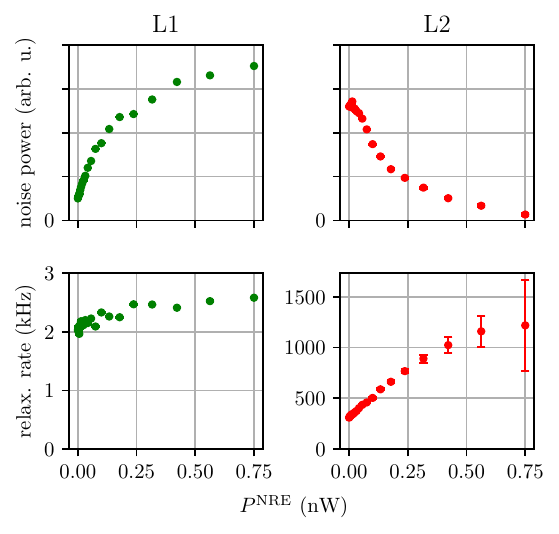}
    \caption{\label{fig:SN_int_dep_ABE}
        Parameters of the slow (L1) and fast (L2) Lorentzian components of individual spin noise spectra as a function of \NREP.
        The noise amplitude of L1 increases significantly with \NREP, indicating a higher number of switching events associated with Si donors. This is accompanied by an increase in the corresponding relaxation rate. In contrast, the noise power of L2 decreases markedly, suggesting a reduction in hole recharging events, while its relaxation rate also increases. Together, these trends suggest a scenario of faster hole charging accompanied by an increased dwell time of the hole in the QD.
    }
\end{figure}

Figure~\ref{fig:SN_typical_LL} displays a representative spin noise (SN) spectrum near the QD resonance, revealing two components, each described by a Lorentzian contribution: a slow component (L1) in the kilohertz range, attributed to the charge dynamics of the Si donors, and a fast component (L2) in the 100\,kHz range, associated with the tunneling dynamics of the resident hole in the QD for the $X^+$ state. We want to emphasize that this assignment was only possible with the combined insights gained from the RF and SNS measurements.

% The SNS measurements also reveal that not only does the hole recharging of the QD speed up with increasing NRE power, as shown by RF in Fig.~\ref{fig:X+_ABE_sat}, but also that the tunneling rate out of the QD decreases with increasing NRE power.

Introducing a second, non-resonant laser with a higher photon energy results in an increase in the rates of both components, as illustrated in Fig. \ref{fig:SN_int_dep_ABE}. Notably, the noise power associated with L1 rises significantly with \NREP, marking an overall increase in Si donor related switching events, which is consistent with the observations in the RF measurements. In contrast, the amplitude of L2 decreases, suggesting a reduction in the number of hole tunneling events. Interestingly, despite the reduced amplitude, the rate associated with L2 increases. This apparent contradiction can be explained by an enhanced hole tunneling rate into the QD, accompanied by an increased residence time of the hole within the dot. We attribute this behavior to carbon impurities in the vicinity of the QD becoming saturated with holes due to the non-resonant excitation.

%%%%%%%%%%%%%%%%%%%%%%%%%%%%%%%%%%%%%%%%%%%%%%%%%%%%%%%%%%%%%%%%%%%%%%%%%%%%%%%%%%%%%%%%%%%%%%%%%%%%%%%%%%%%
\section{Conclusion}\label{sec:conclusion}
This study examined a single GaAs QD in terms of its impurity environment, identifying several adjacent impurity sites that cause spectral shifts of the QD transition. The time-resolved, single-photon resonance fluorescence measurements not only show the typical jumps of tenths of \unit{\micro\eV} for the QD transitions but also identify shifts that are significantly smaller than the QD's homogeneous linewidth and occur rarely. These small shifts are not resolved in standard RF measurements but influence the fidelity of single-photon sources based on QDs. The measurements suggest that the shifts result from charge fluctuations of Si donors, which are close to the Si-doped n-contact. Avoiding these Si donors is challenging since the diffusion constant of Si in (AlGa)As is significantly larger than that for C~\cite{chen_si_1994, beernink_si_1995}. An effective solution might involve slightly increasing the Al fraction at the interface transition from Si-doped to undoped (AlGa)As. This adjustment would elevate the energy levels of the detrimental Si impurities, reducing the probability of charge fluctuations. However, care must be taken not to incorporate DX centers with energy levels below the conduction band by increasing the Al-concentration~\cite{mooney_deep_1990, munoz_techniques_1993}.

Gate voltage dependent RF measurements enabled a detailed study of the QD at various charge states. For negative voltages, the RF shows a weak $X^+$ transition, with count rates more than an order of magnitude lower than those of the $X$ and $X^-$ transitions. However, the mechanism by which the QD is charged with a hole remains unclear in the case of $X^+$. Hole transport from the p-contact to the QD should not be feasible due to the low sample temperature and the built-in, counteracting electric field. Other experiments have observed the $X^+$ transition before in similar structures~\cite{zhai_low-noise_2020}. Here, the origin of the QD hole occupation has been assigned, inter alia, to unintentional C doping in the past~\cite{hayne_optically_2004}, which is unavoidable in molecular beam epitaxy. However, according to the band structure, all carbon acceptors in the regime between the n-contact and the QD should be, in very good approximation, ionized.
The creation of holes in the (AlGa)As by below bandgap light due to two-photon absorption or the photo-assisted Shockley-Read-Hall mechanism \cite{vest_competition_2016} seems reasonable, but the RF photon rate of the $X^+$ transition is, in our experiment, independent of below bandgap optical excitation, ruling out these mechanisms.
Quasi-resonant absorption of the RF laser, probing the $X^+$ transition, could theoretically occur at the nearby $X$ transition—relevant when $X^+$ loses a hole—if the excited exciton's electron simultaneously tunnels out of the QD. However, in our experiment, the RF photon rate increases approximately linearly with the probe laser intensity, which does not support this mechanism. Therefore, the specific physical mechanism responsible for loading the QD with a hole remains an open question.

The experiment clearly shows that the $X^+$ transition can be efficiently recharged through non-resonant optical excitation of the QD quasi-continuum. In this process, the electron from the excited exciton rapidly tunnels out of the QD, leaving a single hole behind. Interestingly, a fast recharging of the QD also increases the dwell time of the hole in the QD, indicating that the hole tunneling out of the QD occurs by tunneling into ionized acceptor states and not into the (AlGa)As continuum.

A lineshape analysis of the $X$, $X^-$, and $X^{2-}$ transitions revealed that, for our low-noise PIN structure, there are very good approximations of purely Lorentzian lineshapes for the $X^-$ and $X^{2-}$ transitions, and a mixture of Lorentzian and Gaussian lineshapes for the $X$ transition.
This is interesting in view of transform-limited single-photon sources and, in the case of $X^-$, with respect to potential spin-photon interfaces. However, one must keep in mind that $X^-$ and $X^{2-}$ are subject to Auger recombination, which has a low, yet non-zero probability in GaAs QDs.

\section{Acknowledgements}
We thank Ronny H\"uther for technical assistance and Nadine Viteritti for processing the sample. This work was funded by the Deutsche Forschungsgemeinschaft (DFG, German Research Foundation) under Germany's Excellence Strategy -- EXC-2123 QuantumFrontiers -- 390837967 and OE 177/10-2.

\appendix

\bibliography{references}

\end{document}